# Role of MgO impurity on the superconducting properties of $MgB_2$

Dharmendra Kumar Singh, Brajesh Tiwari, Rajveer Jha, H. Kishan and V. P. S. Awana[*]

CSIR- National Physical Laboratory, Dr. K.S. Krishnan Marg, New Delhi-110012, India

**Abstract:**

We address the effect of MgO impurity on the superconducting properties of $MgB_2$. The synthesis of $MgB_2$ is very crucial because of sensitivity of Mg to oxidation which may lead to MgO as a secondary phase. Rietveld refinement was performed to determine the quantitative volume fraction of MgO in the samples synthesized by two different techniques. Both the samples were subjected to magnetization measurements under dc and ac applied magnetic fields and the observed results were compared as a function of temperature. Paramagnetic Meissner effect has been observed in a sample of $MgB_2$ having more amount of MgO (with $T_c$ = 37.1K) whereas the pure sample $MgB_2$ having minor quantity of MgO shows diamagnetic Meissner effect with $T_c$ = 38.8K. M-H measurements at 10K reveal a slight difference in irreversibility field which is due to MgO impurity along with wide transition observed from ac magnetic susceptibility measurements. The magnetotransport measurements ρ(T) using $ρ_N$ = 90%, 50% and 10% criterion on pure sample of $MgB_2$ has been used to determine the upper critical field whereas the sample having large quantity of MgO does not allow these measurements due to its high resistance.

[*]**Corresponding Author**
Dr. V. P. S. Awana, Senior Scientist
E-mail: awana@mail.npindia.org
Ph. +91-11-45609357, Fax-+91-11-45609310
Homepage www.freewebs.com/vpsawana/

**Introduction:**

The discovery of superconductivity in $MgB_2$, a simple binary compound in the year 2001, with $T_c$ = 39K [1] has been a very active research area. $MgB_2$ has high $T_c$, lower anisotropy, larger coherence length and transparency of the grain boundaries to current flow making it a very promising candidate for engineering and technological applications [2]. But the critical parameters such as low critical current density ($J_c$), lower upper critical field ($H_{c2}$) and irreversible field ($H_{irr}$) of pure $MgB_2$ limit its potential for industrial applications. Since long several efforts have been made to improve these critical parameters of $MgB_2$ [3-9]. There are two bands in $MgB_2$ (sigma and Pi) with three scattering channels (two intra-bands and one inter-band) having different scattering rates. The improvement in $H_{c2}$ value [3-4] can be achieved by tuning the scattering channels through controlled chemical doping. The critical current density and irreversible field of pure $MgB_2$ have been improved and reported by pinning its vortices by additives of suitable nano-particles such as nano-SiC [5-7], carbon nanotubes (CNT) [8] and nano-diamond [9].

Till date various techniques have been developed for the synthesis of phase pure $MgB_2$ such as high pressure synthesis [10], hybrid microwave [11], self-propagating high temperature synthesis [12], low-temperature solid state reaction [13] and so on. However, to synthesize a phase pure $MgB_2$ remains still a great challenge because firstly of volatile nature of Mg and secondly its prone to oxidation. Hence superconducting properties of $MgB_2$ are very sensitive to the synthesis parameters. In the present study we synthesized two sets of $MgB_2$ samples under different conditions resulting in dramatically different physical properties due to the presence of hidden MgO. The percentage of MgO`s presence in the synthesized $MgB_2$ samples and their effect on magneto-transport properties are reported.

**Experimental details:**

Two polycrystalline bulk $MgB_2$ samples were synthesized by solid-state reaction route. For the synthesis of the samples, Mg powder (from reidel-de-Haen of assay 99%) and amorphous boron powder (from Flukaof assay 95-97%) were taken in 1:2 ratio; the first sample was grounded in ambient atmosphere while the second sample was mixed in a glove box containing argon gas. Both the samples were then pelletized using hydraulic press. The pellets were sealed in vacuum encapsulated quartz tube and sintered in furnace at 750$^o$C for 2.5 hours followed by normal cooling to room temperature.

The diffraction pattern of the synthesized samples was recorded using Rigaku Miniflex II powder x-ray diffractometer (CuK$_α$: λ=1.54 Å). Rietveld refinement was perfomed using FullProf software [14] to analyze the phase purity of samples. Microstructure analysis was also carried out using Field Emission Scanning Electron Microscopy (FE-SEM). Quantum Design`s Physical Property Measurement System (PPMS) was used for the magnetization measurements. Standard four-probe technique was employed for the magneto-resistivity measurements.

**Results and Discussion:**

Figures 1(a) and 1(b) display the XRD patterns recorded at room temperature of two differently prepared samples, one grounded in air (say S1) while another in glove box having Ar atmosphere (say S2). The expected impurity phase in MgB$_2$ is MgO which is due to the oxidation of metallic Mg [15,16]. Both the samples appeared to be same as pure MgB$_2$ in the first look with a trace of MgO at 2θ ≈ 63$^o$, a reflection corresponds to (220) plane of MgO. In order to quantify the volume fraction of MgO in the samples of MgB$_2$, the observed XRD patterns are Reitveld multiphase refined in samples of S1 and S2. The generated patterns (red circles) in space group P6/mmm (#191) and MgO with space group Fm-3m (# 225) are refined together to obtain the minimum difference (blue line) as shown in Figure 1. Bragg reflections in fig.1 for MgB$_2$ are presented by red bars while it is blue for MgO. Multi-phase Rietveld refinement of both the samples revealed that the sample S1 is having larger peak of MgO at 2θ≈ 43$^o$ [inset of fig. 1(a)] compared to the sample S2 as can be seen in the inset of fig. 1(b). This MgO peak at 2θ ≈ 43$^o$ is hidden by the strongest reflection of (101) plane of MgB$_2$. Table-1 depicts the fitness parameter ($χ^2$) of the Reitveld refinement, lattice parameters and quantitative volume fraction of MgO in both the samples. It is clearly seen from this table that the sample S1 contains nearly 39.6% of MgO volume fraction, which is due to oxidation of magnesium, while it is just 7.3% in Sample S2. The lattice parameters do not change significantly in both the samples. However, due to the large presence of MgO in S1, the magnetic and transport properties are quite different from the sample S2 and that is the subject matter of interest in this work.

Scanning electron micrographs are recorded on compact pellets of the samples in order to further confirm the presence of MgO. Figures 2(a) and 2(b) show the SEM images of sample S1 and S2 respectively. The bright patches represent the insulator (MgO) where as dark area shows the conducting MgB$_2$ compound in both the samples. There is a larger distribution of MgO in sample S1 compared to sample S2 ensuring the presence of larger amount of MgO in the sample S1, which is also confirmed by the observation of multi-phase Rietveld refinement of XRD.

Figures 3(a) and 3(b) show the measurements of dc magnetization (M) versus temperature for sample S1 (H=50 Oe) and S2 (H=10 Oe) under zero field cooling (ZFC) and field cooling (FC) situations. A broad superconducting transition (diamagnetic) at around 37.1K is observed, which increases as the temperature decreases, for the sample S1 from the ZFC curve [Fig. 3(a)] indicating the in homogeneity in the sample. For the sample S2 there is a sharp diamagnetic transition at 38.8K, which remains almost constant in the temperature range of 35K to 20 K as can be seen in figure 3(b). Field Cooled magnetization measurements exhibit diamagnetic Meissner effect for sample S2 while an unusual behavior called paramagnetic Meissner effect at around 37.1K for sample S1. The observed behavior of paramagnetic Meissner effect in S1 can be understood as mesoscopic defect structures in $MgB_2$, which is similar to that as reported by Braunisch et al. for Bi-2212 superconductor [17,18] and not related to paramagnetic or ferromagnetic impurity. This observation can be manifestation of weak links caused by insulator MgO defect layers sandwiched between $MgB_2$ grains.

Magnetization measurements of an alternating magnetic field of varying amplitudes say 3Oe, 7Oe and 11Oe at a fixed frequency of f= 333 Hz in the absence of dc bias have been carried out on both the samples. Figures 4(a) and 4(b) show the observed real part (M′) and the imaginary part (M″) of a.c. magnetization for both the samples respectively. M′ for sample S1 shows a broad transition similar to dc magnetization while there is a sharp drop in M′ value for sample S2 below its $T_c$. Moreover, M′ of S1 is weaker in comparison to the sample S2 implying poor diamagnetic shielding in sample S1 due to large volume fraction of MgO. In both the samples below their $T_c$, a peak is observed in M″ measurements at each alternating fields, reflecting intragranular losses due to flux penetrating inside the grains. At the same time the peaks are sharp in sample S2 compared to sample S1 implying homogeneity in the sample S2. It is also important to note that there is a large asymmetry in M″ peaks for sample S1 suggesting inhomogeneous flux distribution in $MgB_2$ due to the presence of large volume of MgO. No evidence of granularity or intergrain coupling is present in both the samples as only one peak is observed even at different applied alternating fields. In addition to this in sample S1, there is a shift in peak position of M″ towards the lower temperature with the increase of amplitude of ac field whereas it is almost constant in case of sample S2 (pure $MgB_2$).

Magnetization measurements at 10 K have been carried out as a function of applied field in order to investigate further the difference in magnetic properties. Figures 5(a) and 5(b) show the M-H loops for the samples S1 and S2 respectively. The irreversibility field ($H_{irr}$), the field at which pinning forces becomes zero, is an important critical parameter from application point of view. The $H_{irr}$ is above 6 T for the sample S2 which is reduced slightly to 5.5T for the sample S1. The observed magnetization value is fifty times higher for the sample S2 (pure $MgB_2$) at low applied fields as compared to the sample S1, which is possibly due to less volume fraction of MgO in S2. Fluxoid jumps are also seen in S2 similar to that as reported in ref [19] while there is no fluxoid jumps in the sample S1. The increase in volume fraction of MgO in the $MgB_2$ (S1) may be responsible to drop off the critical parameters of $MgB_2$ that is also reflected in observing the decrease in M-H loop signal and irreversible field value.

Large presence of insulating MgO in $MgB_2$ makes sample S1 too insulating to carry out the resistivity measurement as a function of temperature. Thus resistivity versus temperature ($\rho$-T) is measured only for the sample S2 under various magnetic fields up to 14 T as shown in figure 6. A very sharp superconducting transition ($\rho \rightarrow 0$) at 38.8 K in absence of external field with transition width ($\Delta T = T_c^{onset} - T_c^{offset}$) of 0.5 K has been observed for the sample S2. The superconducting transition temperature decreases in conjugation with increase in transition width as applied magnetic field increases. The $T_c$ is recorded as 7.7 K at 14 T magnetic field with transition width of 13.5 K as shown in figure 6. The magnetotransport characteristics of sample S1 could not be performed because the sample was showing very high resistance implying degradation of grain connectivity. It seems that the MgO layer is present between $MgB_2$ grains which forbids the direct connectivity between conducting $MgB_2$ grains. However, dc magnetization of Sample S1 is showing superconducting diamagnetic transition at 37.1 K, though not confirmed by $\rho \rightarrow T$ measurements. Such materials with superconducting-insulating-superconducting junctions in bulk $MgB_2$ having higher volume fraction of MgO can be used for microwave source and receiver [20]. A more detailed investigation is needed to ensure these applications in superconducting properties of $MgB_2$ with higher content of MgO.

**Conclusion:**

In brief we have presented the synthesis of two samples of $MgB_2$ along with their lattice parameters, magnetization studies and magnetotransport properties. XRD patterns are reitveld fitted taking into account for MgO as secondary phase in the studied specimen. Sample containing larger volume fraction of MgO (S1), shows wide transition at around 37.1 K with paramagnetic Meissner effect whereas the sample containing smaller volume fraction of MgO (S2) shows sharp transition at around 38.8 K with a clear diamagnetic Meissner effect. Diamagnetic shielding of sample S1 is weaker compared to pure sample of $MgB_2$. Further fluxoid jumps is absent in S1 which is observed in pure sample with large magnetization. Sample S2 confirms superconducting transition at 38.8 K at zero field while the upper critical field ($H_{c2}$) estimated by magnetotransport measurements is 21 Tesla.

**Acknowledgments:**

The authors would like to thank their Director Professor R C Budhani for his support and encouragement in the present work. D. K Singh would like to thank the AcSIR for providing the Trainee scientist scholarship for his M.Tech work. R. Jha would like to thank CSIR for Junior Research Fellowship and H. Kishan for emeritus scientist fellowship. This work is supported by DAE-SRC outstanding investigator award scheme to work on search for new superconductor.

**Figure caption:**

Figure 1: (a) Powder XRD pattern for $MgB_2$ sample S1. The inset shows pooled $MgB_2$ and MgO peaks at $2\theta \approx 43°$. (b) Powder XRD pattern for $MgB_2$ sample S2. The inset shows pooled $MgB_2$ and MgO peaks at $2\theta \approx 43°$.

Figure 2: (a) Scanning electron microscope (SEM) image of $MgB_2$ sample S1 and (b) sample S2.

Figure 3: (a) Plot of dc magnetization versus temperature (M-T) in both zero field-cooled (ZFC) and field-cooled (FC) situations at H = 50Oe for $MgB_2$ sample S1. (b) Plot of dc magnetization versus temperature (M-T) in both zero field-cooled (ZFC) and field-cooled (FC) situations at H = 10 Oe for $MgB_2$ sample S2.

Figure 4: (a) a.c. magnetization (Real) versus temperature (M′-T) plot for $MgB_2$ sample S1 at different fields (3 Oe, 7Oe and 11Oe). The inset shows a.c. magnetization (imaginary) versus temperature (M″-T) plot for $MgB_2$ sample S1 at different fields (3 Oe, 7Oe and 11Oe). (b) for sample S2.

Figure 5: a) Magnetic hysteresis M(H) loops plots at 10 K with applied fields (H) of up to ±7 T for $MgB_2$ sample S1. b) Magnetic hysteresis M(H) loops plots at 10 K with applied fields (H) of up to ±6 T for $MgB_2$ sample S2.

Figure 6: Resistivity versus temperature $\rho(T)$ plots under various applied fields up to 14 T for $MgB_2$ sample S2.

Figure 7: Upper critical field versus temperature $H_{c2}(T)$ plot and fitted to 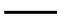 (solid lines) for lower temperature using $\rho_N$ = 90%, 50% and 10% criterion from $\rho(T)$ for $MgB_2$ sample S2.

TABLE 1: Lattice parameters, fitness of Rietveld refinement $\chi^2$, MgO volume fraction of MgB$_2$ (S$_1$) and MgB$_2$ (S$_2$) as calculated from XRD patterns.

|  | Sample-1 (S1) | Sample-2 (S2) |
|---|---|---|
| $\chi^2$ | 6.41 | 4.61 |
| MgO % | 39.63% | 7.23% |
| Lattice parameter of MgB$_2$ | a = b = 3.084(1) A˚, c = 3.530(2) A˚ <br> α = β = 90°, γ = 120° | a = b = 3.086(1) A˚, c = 3.529(2) A˚ <br> α = β = 90°, γ = 120° |
| Lattice parameter of MgO | a = b = c = 4.223(2) A˚ <br> α = β = γ = 90° | a = b = c = 4.223(3) A˚ <br> α = β = γ = 90° |

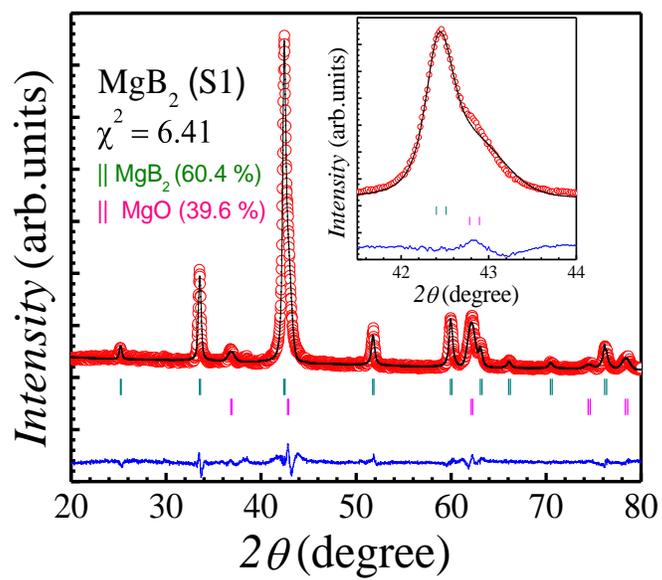

Figure 1(a)

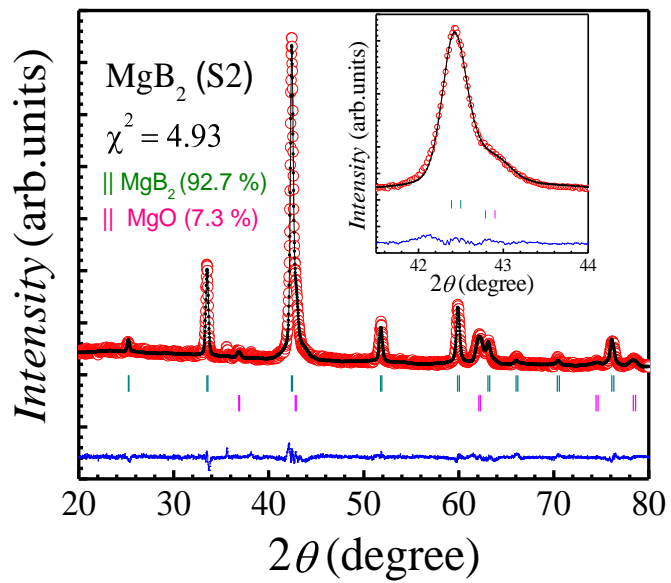

Figure 1(b)

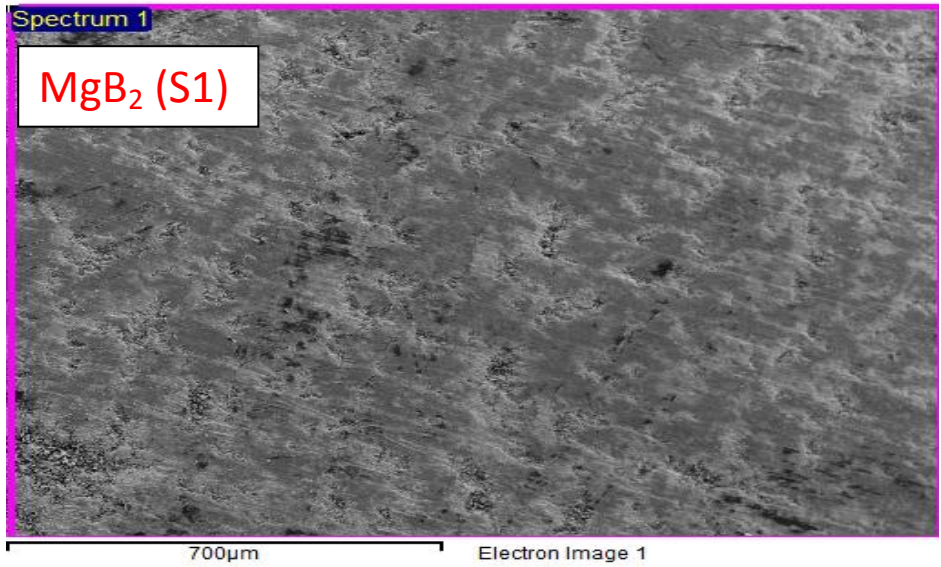

Figure 2(a)

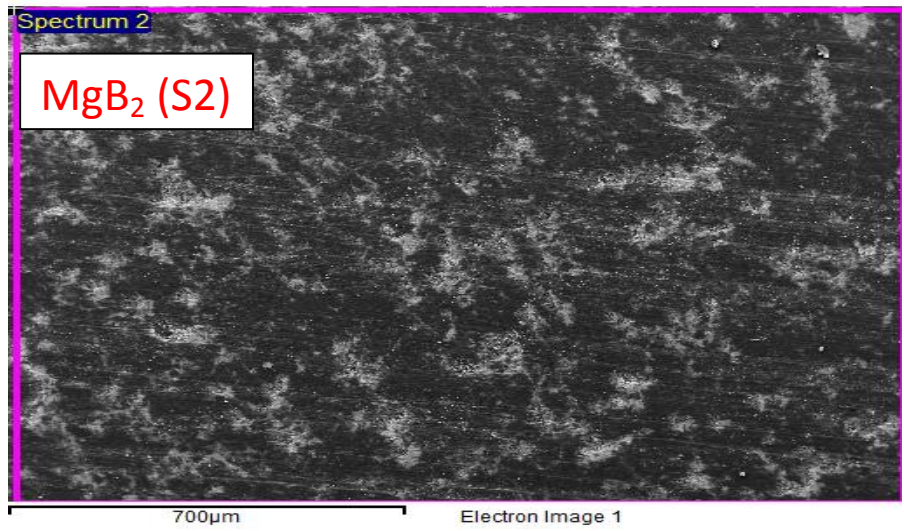

Figure 2(b)

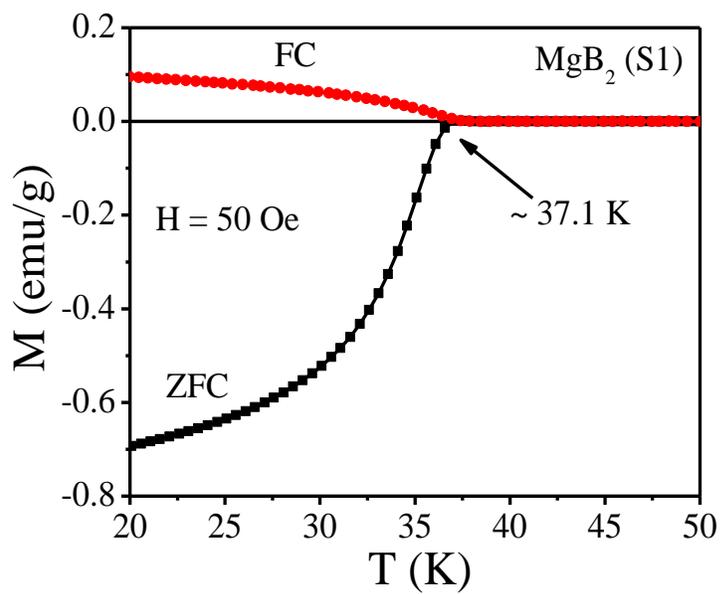

Figure 3(a)

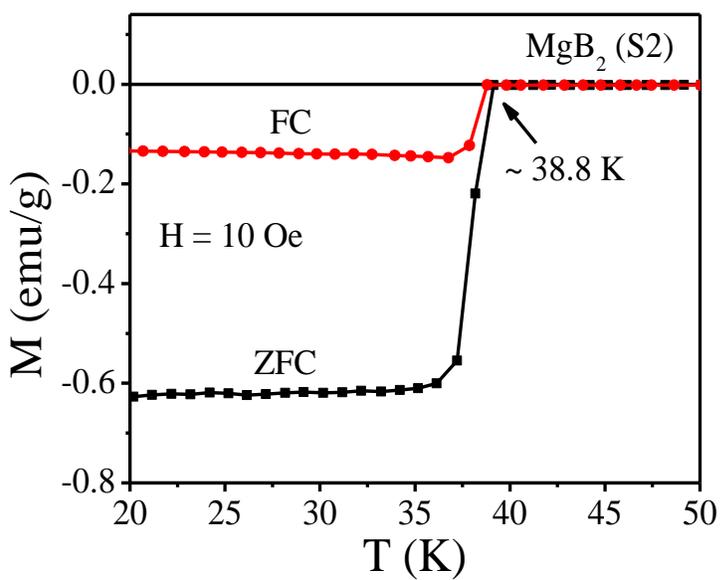

Figure 3(b)

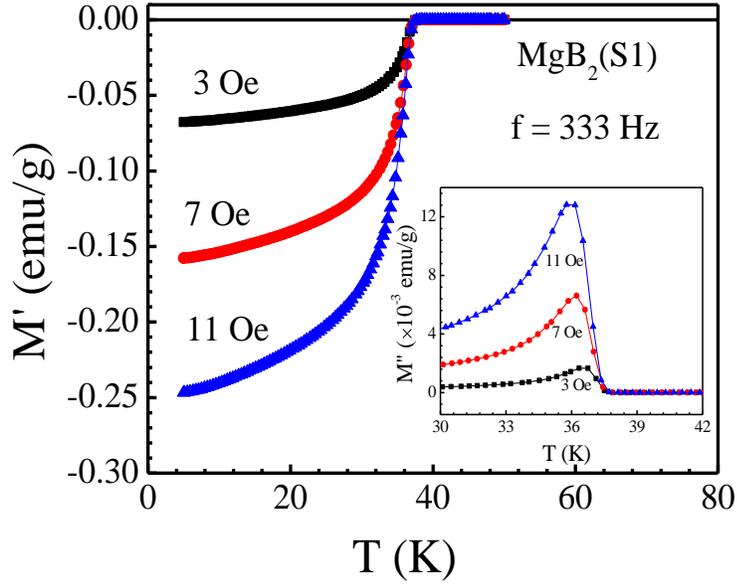

Figure 4(a)

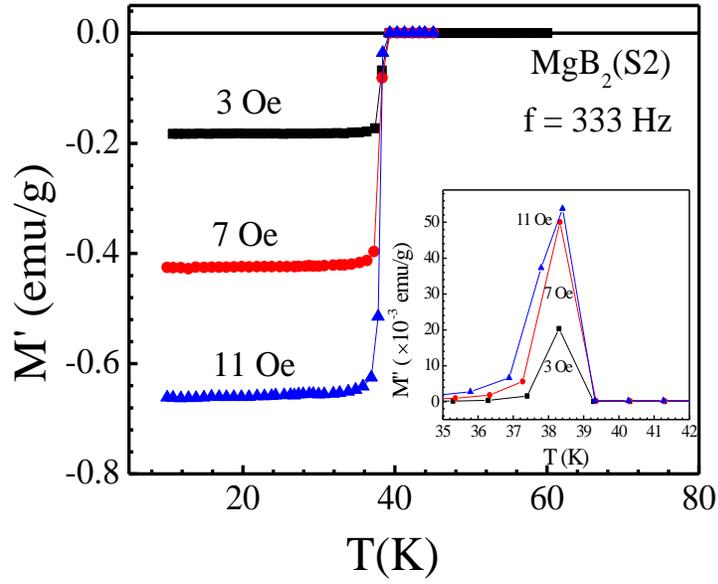

Figure 4(b)

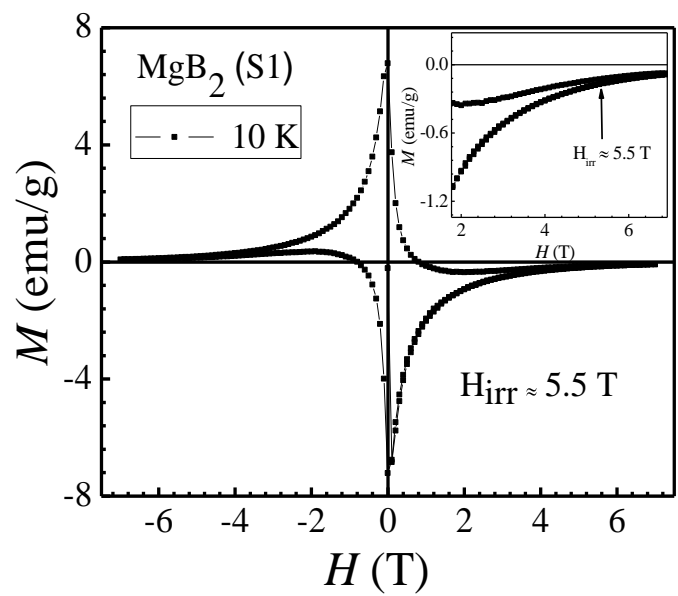

Figure 5(a)

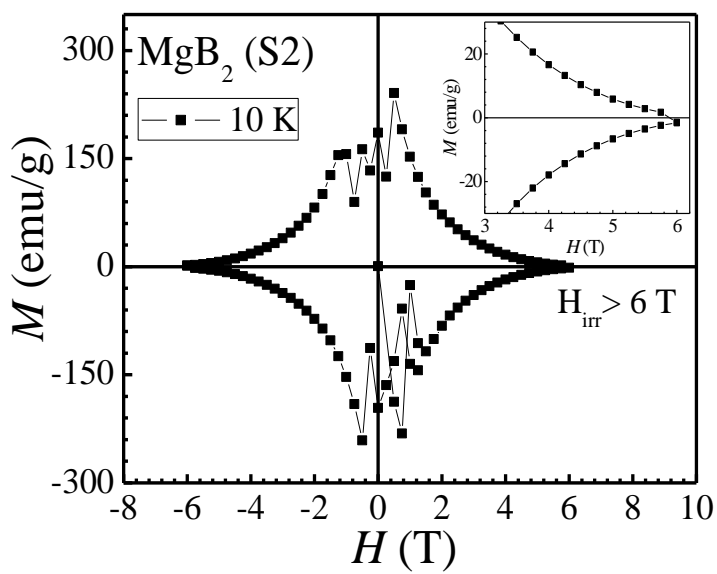

Figure 5(b)

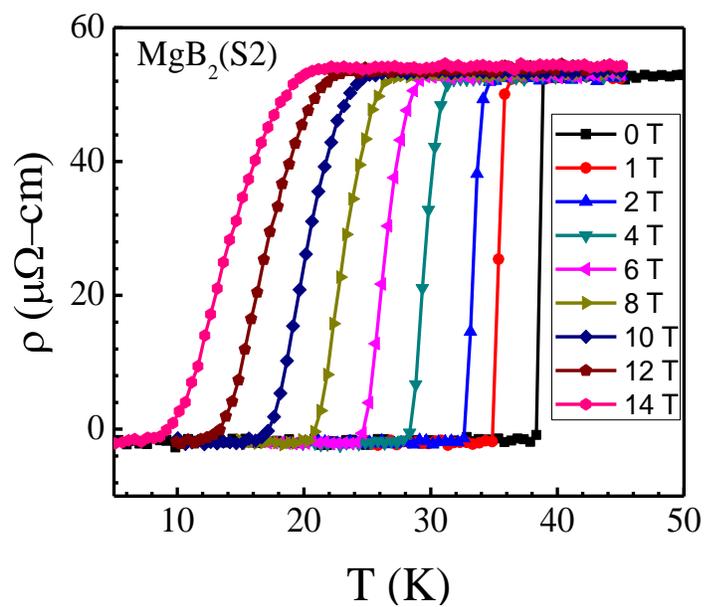

Figure 6

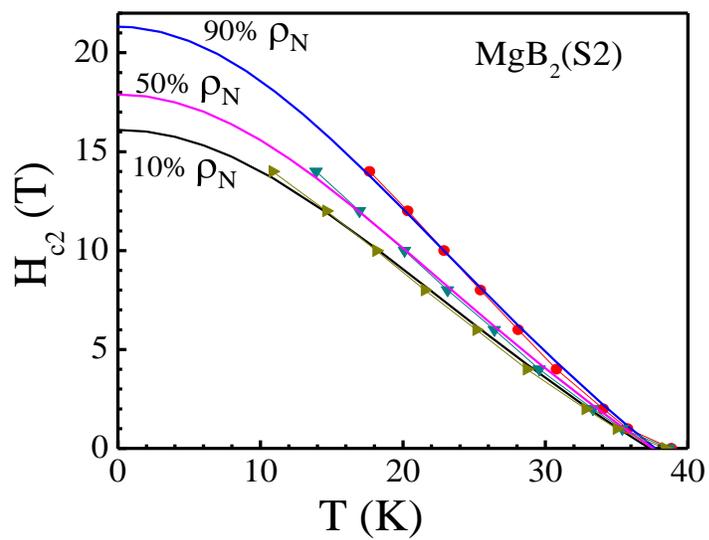

Figure 7

15